\documentclass[pre,amsmath,amssymb,twocolumn,superscriptaddress,showpacs]{revtex4-1}

\pdfoutput=1 

\usepackage{graphicx}
\usepackage{dcolumn}
\usepackage{bm}

\begin{document}

\newcommand{\bea}{\begin{eqnarray}}
\newcommand{\eea}{  \end{eqnarray}}
\newcommand{\bit}{\begin{itemize}}
\newcommand{\eit}{  \end{itemize}}

\newcommand{\be}{\begin{equation}}
\newcommand{\ee}{\end{equation}}
\newcommand{\ra}{\rangle}
\newcommand{\la}{\langle}
\newcommand{\U}{\widetilde{U}} 
\newcommand{\dpc}[1]{\textcolor{red}{(DP: #1)}}
\newcommand{\ntxt}[1]{\textcolor{blue}{#1}} 
\newcommand{\rhop}{\hat{\rho}} 
\newcommand{\he}{\hbar_{\rm eff}}   


\def\bra#1{{\langle#1|}}
\def\ket#1{{|#1\rangle}}
\def\bracket#1#2{{\langle#1|#2\rangle}}
\def\inner#1#2{{\langle#1|#2\rangle}}
\def\expect#1{{\langle#1\rangle}}
\def\e{{\rm e}}
\def\proj{{\hat{\cal P}}}
\def\tr{{\rm Tr}}
\def\H{{\hat H}}
\def\Hdag{{\hat H}^\dagger}
\def\Lop{{\cal L}}
\def\Ehat{{\hat E}}
\def\Edag{{\hat E}^\dagger}
\def\Shat{\hat{S}}
\def\Sdag{{\hat S}^\dagger}
\def\Ahat{{\hat A}}
\def\Adag{{\hat A}^\dagger}
\def\U{{\hat U}}
\def\Udag{{\hat U}^\dagger}
\def\Zhat{{\hat Z}}
\def\Phat{{\hat P}}
\def\Op{{\hat O}}
\def\id{{\hat I}}
\def\x{{\hat x}}
\def\P{{\hat P}}
\def\Px{\proj_x}
\def\Pr{\proj_{R}}
\def\Pl{\proj_{L}}


\title{Chaos simplifies quantum friction}

\author{Gabriel G. Carlo}
\email{carlo@tandar.cnea.gov.ar}
\affiliation{Departamento de F\'\i sica, CNEA, CONICET, Libertador 8250, (C1429BNP) Buenos Aires, Argentina}  
\author{Leonardo Ermann}
\affiliation{Departamento de F\'\i sica, CNEA, CONICET, Libertador 8250, (C1429BNP) Buenos Aires, Argentina}
\author{Alejandro M. F. Rivas}
\affiliation{Departamento de F\'\i sica, CNEA, CONICET, Libertador 8250, (C1429BNP) Buenos Aires, Argentina}

\date{\today}

\pacs{05.45.Mt, 03.65.Yz, 05.45.−a}
  
\begin{abstract}

By means of studying the evolution equation for the Wigner distributions of quantum dissipative systems we derive 
the quantum corrections to the classical Liouville dynamics, taking into account the standard quantum 
friction model. The resulting evolution turns out to be the classical one plus fluctuations that depend not 
only on the $\hbar$ size but also on the momentum and the dissipation parameter (i.e. the 
coupling with the environment). On the other hand, we extend our studies of a paradigmatic 
system based on the kicked rotator, and we confirm that by adding fluctuations only depending on the size 
of the Planck constant we essentially recover the quantum behaviour. This is systematically measured in the parameter 
space with the overlaps and differences in the dispersion of the marginal distributions 
corresponding to the Wigner functions. Taking into account these results and analyzing the Wigner 
evolution equation we propose that the chaotic nature of our system is responsible for the independence on 
the momentum, while the dependence on the dissipation is provided implicitly by the dynamics.

\end{abstract}

\maketitle

\section{Introduction}
\label{sec1}

There is a long standing connection between quantum chaos and decoherence \cite{SaracenoPaz}, that has proven 
to be very fruitful in discovering quantum to classical correspondence properties. Nowadays we witness a renovated 
interest in complex and dissipative phenomena. For instance the decoherence rate in fluctuating chaotic quantum systems shows 
an exponential behaviour with the particle number, turning them into ideal candidates for wave function collapse models 
\cite{DelCampo}. Moreover, there is an impressive recent amount of work in quantum chaos that is related to a 
measure of dynamical complexity, the out-of-time-ordered correlators \cite{OTOC}, which has its roots in quantum 
field theory and black hole studies \cite{qfield}. 

There are also many other areas that have evolved in recent times and that are intimately related to 
quantum complex dissipative systems. We can mention optomechanics \cite{Bakemeier} and many body systems 
\cite{Hartmann} which could greatly benefit from a better understanding of the quantum consequences 
of classical dissipative chaos. There are promising first attempts to treat quantum bifurcations \cite{Ivanchenko,Wang} that 
could be very important in elucidating the properties of Floquet crystals. Also, the details of the environmental 
models, no matter how subtle, could have very strong effects on the degree of complexity \cite{Eastman}. As a matter 
fact, this idea has been turned into a proposal \cite{Eastman2} in order to 
control chaos at the quantum level. 
Very recently \cite{Bondaretal}, the relevance 
of the quantum friction model details have been analyzed, underlining the need for a better understanding of 
even this fundamental mechanism. This a crucial issue not only in view of all the 
intense research work that is being carried out, but because this kind of dissipation is at the base of any extension of the rich 
dynamical systems theory to the quantum realm.

In order to look into the quantum properties of classically chaotic and dissipative systems 
we have derived the evolution equation for the Wigner distributions up to order $\hbar$. We have considered 
the standard quantum friction model that we used in previous work. As a result we 
identify many ingredients that appear as quantum corrections to the classical evolution for Liouville distributions. 
These are dependent not only on the size of $\hbar$, but also on phase space variables 
like the momentum, and the friction parameter. 
We compare with comprehensive calculations on the whole parameter space corresponding to a paradigmatic 
model of dissipative quantum chaos, i.e. the dissipative modified kicked rotator map (DMKRM) \cite{qdisratchets}. 
We systematically confirm that a very simple model of classical evolution plus Gaussian noise of the size of $\hbar$ captures 
all the main details of the marginal distributions corresponding to the Wigner functions of the equilibrium states. 
This is measured by calculating the overlaps between the quantum and classical distributions and the difference 
in the dispersions, for several finite values of the effective Planck constant $\he$.
We provide with a complete explanation for this simplification that we attribute to the chaotic nature of our system and 
an implicit dependence given by the dynamics itself.

This paper has the following structure: In Sec. \ref{sec2} we derive the evolution equation for the Wigner 
function and specialize it for the case of the standard friction model as the only source of dissipation. 
In Sec. \ref{sec3} we explore  the whole parameter space of the DMKRM by calculating the overlaps between the 
classical and quantum marginal distributions and the difference in the dispersions. We explain 
the robust validity of our very simple classical model as a consequence of chaotic behaviour. 
In Sec. \ref{sec4} we give our conclusions and outlook.

\section{Wigner function dissipative evolution equation. The standard friction model}
\label{sec2}

In classical mechanics, dissipative systems are governed by the Fokker-Planck
equation which describes the evolution of Liouville distributions
in phase space. On the other hand, the evolution of quantum dissipative systems
is governed by the Lindblad equation, which expresses the evolution
of density operators in their corresponding Hilbert space. 
In order to perform an explicit comparison between quantum and classical
mechanics, it is crucial to express both in the same framework. 

The Weyl-Wigner representation provides with a phase space description
of quantum mechanics \cite{Curtright1998,Curtright2011,Curtright2013},
where an observable $O=O(q,p)$ is assumed to be a real-valued function
of coordinate $q$ and momentum $p$, and the state of the system is represented
by the Wigner function $W=W(q,p)$. Wigner functions provide with a very 
suitable quantum analogue of Liouville distributions. 
Recently, using the Weyl-Wigner formalism and
the star product, Bondar et al. \cite{Bondaretal} have obtained 
the Lindblad-Wigner equation which describes the evolution of the
Wigner functions of dissipative systems in phase space. 
We closely follow their approach and consider 
\begin{equation}
\frac{d}{dt}W=-\frac{i}{\hbar}(H\star W-W\star H)+D[W]+D'[W], 
\label{eq:WLeq}
\end{equation}
where the first term accounts for the evolution given by the Hamiltonian 
\[
H=p^{2}/(2m)+U(q), 
\]
with $m$ the mass of the particle and $U(q)$ the potential. 
The second term in Eq. (\ref{eq:WLeq}), the dissipator, involves the Weyl-Wigner 
representation of the Lindblad operators $L$ and is given by
\[
D[W]=\frac{2\nu}{\hbar}\Big(L\star W\star L^{*}-\frac{1}{2}W\star L^{*}\star L-\frac{1}{2}L^{*}\star L\star W\Big), 
\]
where $\nu$ is the dissipation parameter. The last term is   
\[
D'[W]=\frac{2\mathfrak{D}}{\hbar^{2}}\Big(q\star W\star q-\frac{1}{2}W\star q\star q-
\frac{1}{2}q\star q\star W\Big)=\mathfrak{D}\frac{\partial^{2}W}{\partial p^{2}}
\]
and accounts for thermal diffusion of size $\mathfrak{D}$ 
(which we do not take into account in the present work).
The star product between two phase space distributions $A$ and $B$ is defined as 
\[
A\star B=A\exp\frac{i\hbar}{2}\left(\overleftarrow{\frac{\partial}{\partial q}}
\overrightarrow{\frac{\partial}{\partial p}}-\overleftarrow{\frac{\partial}{\partial p}}
\overrightarrow{\frac{\partial}{\partial q}}\right)B=A\left[\exp\frac{i\hbar}{2}\overleftrightarrow{\nabla}\right]B,
\]
and can be expressed in terms of the symplectic differential product $\overleftrightarrow{\nabla}$ 
\begin{equation}
\overleftrightarrow{\nabla}=\left(\overleftarrow{\frac{\partial}{\partial q}}\overrightarrow{\frac{\partial}{\partial p}}-
\overleftarrow{\frac{\partial}{\partial p}}\overrightarrow{\frac{\partial}{\partial q}}\right)=
\overleftarrow{\nabla}\wedge\overrightarrow{\nabla}=
\overleftarrow{\frac{\partial}{\partial x_{i}}}J_{ij}\overrightarrow{\frac{\partial}{\partial x_{j}}},
\label{eq:DoubleDelta}
\end{equation}
where $x_1=q$ and $x_2=p$, and the symplectic matrix $J$ is given by 
\begin{equation}
J=\left[\begin{array}{cc}
0 & 1\\
-1 & 0
\end{array}\right].\label{eq:Jxp}
\end{equation}
In fact, 
\[
A\overleftrightarrow{\nabla}B=\frac{\partial A}{\partial x_{i}}J_{ij}\frac{\partial B}{\partial x_{i}}=
\frac{\partial A}{\partial q}\frac{\partial B}{\partial p}-\frac{\partial A}{\partial p}\frac{\partial B}{\partial q}, 
\]
and then the star product expansion in powers of $\hbar$ has the following compact shape 
\[
A\star B=AB+\frac{i\hbar}{2}A\overleftrightarrow{\nabla}B-\frac{\hbar^{2}}{8}A\overleftrightarrow{\nabla^{2}}B+O(\hbar^{3}),
\]
where the double differential operator is simply  
\begin{equation}
A\overleftrightarrow{\nabla^{2}}B=
\frac{\partial^{2}A}{\partial x_{i}\partial x_{k}}J_{ij}J_{kl}\frac{\partial^{2}B}{\partial x_{j}\partial x_{l}}.
\label{eq:Deltacuadrado}
\end{equation}

Performing the $\hbar$ expansion for the Hamiltonian term in Eq. (\ref{eq:WLeq}), we obtain 
\begin{eqnarray}
-\frac{i}{\hbar}(H\star W-W\star H)& =& -\frac{i}{\hbar}\left[(HW-WH) \nonumber \right. \\
&+& \left. \frac{i\hbar}{2}(H\overleftrightarrow{\nabla}W-W\overleftrightarrow{\nabla}H) \nonumber \right. \\
&+& \left. \frac{1}{2}\left(\frac{i\hbar}{2}\right)^{2}(H\overleftrightarrow{\nabla^{2}}W-
W\overleftrightarrow{\nabla^{2}}H) ]. \nonumber \right. \\ 
& \ & \left. \ \right.
\end{eqnarray}
Using the definitions of Eq. (\ref{eq:DoubleDelta}) and Eq. (\ref{eq:Deltacuadrado})
we get 
\begin{eqnarray}
-\frac{i}{\hbar}(H\star W-W\star H) &=&\left\{ H,W\right\} -
\frac{i\hbar}{4}\left[\frac{\partial^{2}H}{\partial x^{2}}\frac{\partial^{2}W}{\partial p^{2}} \nonumber \right. \\ 
&+& \frac{\partial^{2}H}{\partial p^{2}}\frac{\partial^{2}W}{\partial x^{2}} 
- \left. 2\frac{\partial^{2}H}{\partial x\partial p}\frac{\partial^{2}W}{\partial p\partial x}\right]. \nonumber \\
\label{qcoarse}
\end{eqnarray}
The first term is exactly the classical evolution determined by the
Poisson bracket $\left\{ H,W\right\} $. The second term, with
an imaginary part even for real valued Hamiltonians and Wigner functions, 
is responsible for the development of the quantum coherences and quantum 
coarse graining.

For the dissipator, expanding it in powers of $\hbar$ we obtain 
\begin{eqnarray}
D[W] & = & \frac{2\nu}{\hbar}\Big(LWL^{*}-\frac{1}{2}WL^{*}L-\frac{1}{2}L^{*}LW\Big)+i\nu S_{1} \nonumber \\ 
 & - & \frac{\nu\hbar}{2}S_{2} = i\nu S_{1}-\frac{\nu\hbar}{2}S_{2}, 
\label{eq:DS1S2}
\end{eqnarray}
 where $S_{1}$ and $S_{2}$ contain terms with simple and double differential product operators respectively. 
 The associative property of the star product
\[
A\star B\star C=A\star(B\star C)=(A\star B)\star C
\]
extends to the different powers in $\hbar$. So that
\begin{eqnarray}
S_{1} & = & L\overleftrightarrow{\nabla}(WL^{*})+L(W)\overleftrightarrow{\nabla}L^{*}
-\frac{1}{2}W\overleftrightarrow{\nabla}(L^{*}L)\nonumber \\
 & - & \frac{1}{2}W(L^{*})\overleftrightarrow{\nabla}L-\frac{1}{2}L^{*}\overleftrightarrow{\nabla}(LW)-
 \frac{1}{2}L^{*}(L)\overleftrightarrow{\nabla}W=\nonumber \\
 &  & L^{*}(L\overleftrightarrow{\nabla}W)+L(W\overleftrightarrow{\nabla}L^{*})+2W(L\overleftrightarrow{\nabla}L^{*}) 
 \label{eq:S1LWL}
\end{eqnarray}
and 
\begin{eqnarray}
S_{2} & = & L\overleftrightarrow{\nabla}(W\overleftrightarrow{\nabla}L^{*})+
\frac{1}{2}L\overleftrightarrow{\nabla^{2}}(WL^{*})+\frac{1}{2}L(W\overleftrightarrow{\nabla^{2}}L^{*})\nonumber \\
 & - & \frac{1}{2}\left[W\overleftrightarrow{\nabla}(L^{*}\overleftrightarrow{\nabla}L)+
 \frac{1}{2}W\overleftrightarrow{\nabla^{2}}(L^{*}L)+\frac{1}{2}W(L^{*}\overleftrightarrow{\nabla^{2}}L)\right]\nonumber \\
 & - & \frac{1}{2} \left[L^{*}\overleftrightarrow{\nabla}(L\overleftrightarrow{\nabla}W)+
 \frac{1}{2}L^{*}\overleftrightarrow{\nabla^{2}}(LW)+\frac{1}{2}L^{*}(L\overleftrightarrow{\nabla^{2}}W) \right]. \nonumber \\
 &  &
 \label{eq:S2LWL}
\end{eqnarray}

We consider Lindblad operators such that their Weyl symbols are of the form, 
\begin{equation}
L(x)=\sqrt{ lf \left( |p| + \frac{\hbar}{2l} \right) } \exp{(-i{\rm sign}{(p)}q/l)},
\label{eq:LindbladOK}
\end{equation} 
where $l$ is a length scale constant and $f$ is an arbitrary function defining the velocity dependence of the 
dissipative force. In \cite{Bondaretal}, it was shown that in these cases the first moments of the Wigner function 
satisfy the Ehrenfest equations characterizing the motion of a particle of mass 
$m$ interacting with an environment induced velocity-dependent friction. 
Also, introducing the expression of Eq. (\ref{eq:LindbladOK}) in Eq. (\ref{eq:S1LWL}) we get
\begin{equation}
S_{1}=-i\frac{\partial}{\partial p}\left[{\rm sign}(p)f\left(\left|p\right|+\frac{\hbar}{2l}\right)W\right].
\label{eq:S1deriv}
\end{equation}
Then, for Lindblad operators whose Weyl symbols obey the form of Eq. (\ref{eq:LindbladOK}), 
the classical limit of the Lindblad-Wigner Eq. (\ref{eq:WLeq}) corresponds to the appropriate Fokker-Planck equation 
\cite{Gardiner1985},
\begin{equation}
D[W]=2\nu\frac{\partial}{\partial p}\left[{\rm sign}(p)f(|p|)W\right]+O\left(\hbar\right).
\label{ClassicalLimit}
\end{equation}

In order to obtain the quantum corrections we need the expression
of $S_{2}$ in Eq. (\ref{eq:S2LWL}). With the help of the symplectic
differential product definitions of Eq. (\ref{eq:DoubleDelta}) and Eq. (\ref{eq:Deltacuadrado}) 
and using the symplectic matrix form of Eq. (\ref{eq:Jxp}) we get, after straightforward calculations, 
\begin{eqnarray*}
S_{2} & = & -2\frac{\partial^{2}W}{\partial x\partial x}\frac{\partial L}{\partial p}\frac{\partial L^{*}}{\partial p}-
2\frac{\partial^{2}W}{\partial p\partial p}\frac{\partial L}{\partial x}\frac{\partial L^{*}}{\partial x}\\
 & + & 2\frac{\partial^{2}W}{\partial x\partial p}\left(\frac{\partial L}{\partial p}\frac{\partial L^{*}}{\partial x}+
 \frac{\partial L}{\partial x}\frac{\partial L^{*}}{\partial p}\right)\\
 & + & \frac{\partial^{2}L^{*}}{\partial x\partial x}\frac{\partial L}{\partial p}\frac{\partial W}{\partial p}+
 \frac{\partial^{2}L^{*}}{\partial p\partial p}\frac{\partial L}{\partial x}\frac{\partial W}{\partial x}\\
 & - & \frac{\partial^{2}L^{*}}{\partial x\partial p}\left(\frac{\partial L}{\partial p}\frac{\partial W}{\partial x}+
 \frac{\partial L}{\partial x}\frac{\partial W}{\partial p}\right)\\
 & + & \frac{\partial^{2}L}{\partial x\partial x}\frac{\partial L^{*}}{\partial p}\frac{\partial W}{\partial p}+
 \frac{\partial^{2}L}{\partial p\partial p}\frac{\partial L^{*}}{\partial x}\frac{\partial W}{\partial x}\\
 & - & \frac{\partial^{2}L}{\partial x\partial p}\left(\frac{\partial L^{*}}{\partial p}\frac{\partial W}{\partial x}+
 \frac{\partial L^{*}}{\partial x}\frac{\partial W}{\partial p}\right).
\end{eqnarray*}
Using  Lindblad operators whose Weyl symbols are of the form given in Eq. (\ref{eq:LindbladOK}), we get 
\begin{eqnarray}
S_{2} & = & -\frac{1}{2}\left(\frac{\partial f}{\partial p}\right)^{2}\frac{l}{f\left(\left|p\right|+
\frac{\hbar}{2l}\right)}\frac{\partial^{2}W}{\partial x\partial x} \nonumber \\
 & - & 2\frac{f\left(\left|p\right|+\frac{\hbar}{2l}\right)}{l}\frac{\partial^{2}W}{\partial p\partial p}-
 \frac{1}{l}\frac{\partial f}{\partial p}\frac{\partial W}{\partial p}.
 \label{eq:S2deriv}
\end{eqnarray}
By employing the expressions of Eq. (\ref{eq:S1deriv}) and Eq. (\ref{eq:S2deriv}) 
for $S_1$ and $S_2$ respectively in the dissipator of Eq. (\ref{eq:DS1S2}) and expanding up to first order in $\hbar$, 
we obtain
\begin{eqnarray*}
D[W] & = & 2\nu\frac{\partial}{\partial p}\left[{\rm sign}(p)f(|p|)W\right]\\
 & + & \frac{\hbar\nu}{l}\left[\frac{3}{2}\frac{\partial f}{\partial p}\frac{\partial W}{\partial p}+
 \frac{\partial^{2}f}{\partial p{}^{2}}W\right]\\
 & + & \hbar\nu\left[\frac{1}{4}\left(\frac{\partial f}{\partial p}\right)^{2}
 \frac{l}{f\left(\left|p\right|\right)}\frac{\partial^{2}W}{\partial x\partial x}+
 \frac{f(|p|)}{l}\frac{\partial^{2}W}{\partial p\partial p}\right]
\end{eqnarray*}
or equivalently, 
\begin{eqnarray}
D[W] & = & 2\nu\frac{\partial}{\partial p}\left[{\rm sign}(p)f(|p|)W\right]\nonumber \\
 & + & \frac{\hbar\nu}{l}\left[\frac{3}{2}\frac{\partial}{\partial p}\left(\frac{\partial f}{\partial p}W\right)-
 \frac{1}{2}\frac{\partial^{2}f}{\partial p{}^{2}}W\right]\nonumber \\
 & + & \hbar\nu\left[\frac{1}{4}\left(\frac{\partial f}{\partial p}\right)^{2}\frac{l}{f\left(\left|p\right|\right)}
 \frac{\partial^{2}W}{\partial x\partial x}+\frac{f(|p|)}{l}\frac{\partial^{2}W}{\partial p\partial p}\right]. \nonumber \\
 &  &
 \label{eq: DWexpand}
\end{eqnarray}
The first term in Eq. (\ref{eq: DWexpand}) gives the
Fokker-Planck part. The second term can be seen as an $\hbar$
correction for the dissipative constant $\nu$, while the last
term performs a $\hbar\nu$ size diffusion on the Wigner function
phase space distribution $W$. Note that no imaginary term is present
in Eq. (\ref{eq: DWexpand}), hence the dissipator does not develop any
coherence. Moreover, coherences will be washed out by the diffusion
implied by the last term.

In our previous studies (\cite{qdisratchets} and subsequent work) we have taken into account Lindblad operators 
$\hat{L}_{\mu}$ given by 
\begin{equation}
\begin{array}{l}
\hat{L}_1 = g \sum_n \sqrt{n+1} \; |n \rangle \, \langle n+1|,\\
\hat{L}_2 = g \sum_n \sqrt{n+1} \; |-n \rangle \, \langle -n-1|,
\end{array}
\end{equation}
where $|n\rangle$ are momentum eigenstates with $n=0,1\ldots$. 
In fact, this is the standard friction model found in the literature \cite{Dittrich, Graham}. 
Identifying $g=\sqrt{2 \nu}$, the dissipator is written in terms of the Lindblad operator 
$\hat{L}=\tilde{L_{1}}+\tilde{L_{2}}$ with
\[
\tilde{L_{1}}=\sqrt{ \frac{\hbar}{2\nu} }
g\sum_{n}\sqrt{n+1}|n\rangle\langle n+1|
\]
and
\[
\tilde{L_{2}}=\sqrt{ \frac{\hbar}{2\nu} }g\sum_{n}\sqrt{n+1}|-n\rangle\langle-n-1|.
\]

Now, for these Lindblad operators we perform the discrete Weyl Wigner transformation in the $N=\frac{1}{2\pi\hbar}$ 
dimensional Hilbert space \cite{opetor}. For the Lindblad operators $\tilde{L_{i}}$ we obtain the Weyl Wigner 
symbols $L_{i}(x)$ as \cite{opetor}
\[
L_{i}(x)=\sum_{k}\langle 2a-k|\tilde{L_{i}|}k\rangle e^{\frac{i2\pi}{N}2(a-k)b}
\]
which are functions of the phase space points  $x=(p,q)=(\frac{a}{N},\frac{b}{N})$, with $a$ and $b$ semi integer numbers.
For $\tilde{L_{1}}$ we have (simplifying the prefactor)
\[
L_{1}(x)=\sqrt{\hbar}\sum_{n}\sqrt{n+1}\sum_{k}\langle 2a-k|n\rangle\langle n+1|k\rangle e^{\frac{i2\pi}{N}2(a-k)b}.
\] 
Performing the summations, with $x=(p,q)=(\frac{a}{N},\frac{b}{N})$ and defining $l=\frac{1}{2\pi}$, we get,
\[
L_{1}(x)=\sqrt{l \left(p+\frac{\hbar}{2l}\right)  }e^{\frac{-iq}{l}}
\]
for positive values of $p$ and zero otherwise.
A similar procedure for $\tilde{L_{2}}$ implies that the Wigner symbol for $\hat{L}$ is 
\begin{equation}
L(x)=     \sqrt{l \left(\left|p\right|+\frac{\hbar}{2l}\right)  }\exp(-i{\rm sign}(p)q/l).
\label{eq:Lx}
\end{equation}
We observe that Eq. (\ref{eq:Lx}) is similar to Eq. (\ref{eq:LindbladOK}) if we take a linear function 
$f\left(\left|p\right|+\frac{\hbar}{2l}\right)=\left|p\right|+\frac{\hbar}{2l}$.
Hence, we verify that the Lindblad operator $\hat{L}=\tilde{L_{1}}+\tilde{L_{2}}$ 
has the correct phase space variables dependence in order 
to satisfy the Ehrenfest equations. For this operator, the evolution of the Wigner function in phase space is 
obtained with the dissipator as in Eq. (\ref{eq: DWexpand}) which for this linear function becomes
\begin{eqnarray}
D[W] & = &  g^2\frac{\partial}{\partial p}\left[{\rm sign}(p)\left(|p|+\frac{3}{4}\frac{\hbar}{l}\right)W\right] \nonumber \\
 & + & g^2\left( \frac{\hbar}{2}\right)\left[\frac{1}{4}\frac{l}{\left|p\right|}\frac{\partial^{2}W}{\partial x\partial x}+
 \frac{|p|}{l}\frac{\partial^{2}W}{\partial p\partial p}\right]. 
 \label{eq: DWlin}
 \end{eqnarray}
This is a classical Fokker-Planck evolution with a dissipation constant $g^2$ slightly corrected with $\hbar$ and a 
diffusion term that scales with $g^2 \hbar$.

\section{Classical and quantum marginal distributions: comparison in a paradigmatic system}
\label{sec3}

\subsection{Dissipative modified kicked rotator map}

For a systematic exploration (in a meaningful region) of the parameter space of a quantum dissipative system 
subject to friction we consider the paradigmatic DMKRM. It corresponds to a particle moving in one dimension
[$q\in(-\infty,+\infty)$] kicked in a periodic fashion by:
\begin{equation}
V(q,t)=k\left[\cos(q)+\frac{a}{2}\cos(2q+\phi)\right]
\sum_{m=-\infty}^{+\infty}\delta(t-m \tau),
\end{equation}
being $k$ the strength of the kick and $\tau$ its period. 
Adding dissipation we get a map with friction as follows \cite{qdisratchets}
\begin{equation}
\left\{
\begin{array}{l}
\overline{n}=\gamma n +
k[\sin(q)+a\sin(2q+\phi)]
\\
\overline{q}=q+ \tau \overline{n}.
\end{array}
\right.
\label{dissmap}
\end{equation}
where $n$ is the momentum variable conjugated to $q$ 
and $\gamma=\exp{-2 \nu}$ ($0\le \gamma \le 1$) is the dissipation parameter.
We recover the conservative system by setting $\gamma=1$, on the opposite side  
$\gamma=0$ corresponds to the maximum friction. One usually introduces a rescaled 
momentum variable $p=\tau n$ and the quantity $K=k \tau$ in order to simplify things. 
This model has been extensively used to study the properties of directed transport, as such 
it shows a net current when breaking the spatial and temporal symmetries (i.e., when $a \neq 0$ with $\phi \neq m
\pi$, and  $\gamma \neq 1$). For historical reasons, we adopt the values $a=0.5$ 
and $\phi=\pi/2$ for the rest of this work.

Inspired in the known general correspondence of the quantum dissipative evolution with roughly a classical evolution with noise, 
the main effects of the quantum fluctuations have been proposed to be similar to those of Gaussian fluctuations of the order of 
$\hbar_{\rm eff}$ imposed on the classical map  \cite{Carlo} (the definition of $\hbar_{\rm eff}$, the effective Planck constant, 
can be found in the next paragraph). 
For that purpose we add $\xi$ (i.e., the random fluctuations) to the first line of Eq. (\ref{dissmap}), 
fixing $\langle \xi^2 \rangle = \hbar_{\rm eff}$, having zero mean. This corresponds to a kind of diffusion of $\he$ size.
This exact identification could be relaxed, since it is enough for the fluctuations to be of the order 
of $\hbar_{\rm eff}$ to comply with this conjecture.

On the quantum side we have that 
$q\to \hat{q}$, $n\to \hat{n}=-i (d/dq)$ ($\hbar=1$).
Given that $[\hat{q},\hat{p}]=i \tau$ (where $\hat{p}=\tau \hat{n}$), we define the effective Planck constant
by means of identifying $\hbar_{\rm eff}=\tau$. In the classical limit 
$\hbar_{\rm eff}\to 0$ and $K=\hbar_{\rm eff} k$ remains constant. 
Dissipation is treated in the usual way with a Lindblad 
master equation \cite{Lindblad} for the density operator $\hat{\rho}$ 
\begin{equation}
\dot{\hat{\rho}} = -i
[\hat{H}_s,\hat{\rho}] - \frac{1}{2} \sum_{\mu=1}^2
\{\hat{L}_{\mu}^{\dag} \hat{L}_{\mu},\hat{\rho}\}+
\sum_{\mu=1}^2 \hat{L}_{\mu} \hat{\rho} \hat{L}_{\mu}^{\dag} \equiv \Lambda \rho.
\label{lindblad}
\end{equation}
$\hat{H}_s=\hat{n}^2/2+V(\hat{q},t)$ is the system
Hamiltonian, \{\,,\,\} the anticommutator, and $\hat{L}_{\mu}$ the Lindblad operators 
defined in Sec. \ref{sec2}.

\subsection{Overlaps and dispersions}

\begin{figure}[htp]
\includegraphics[width=0.47\textwidth]{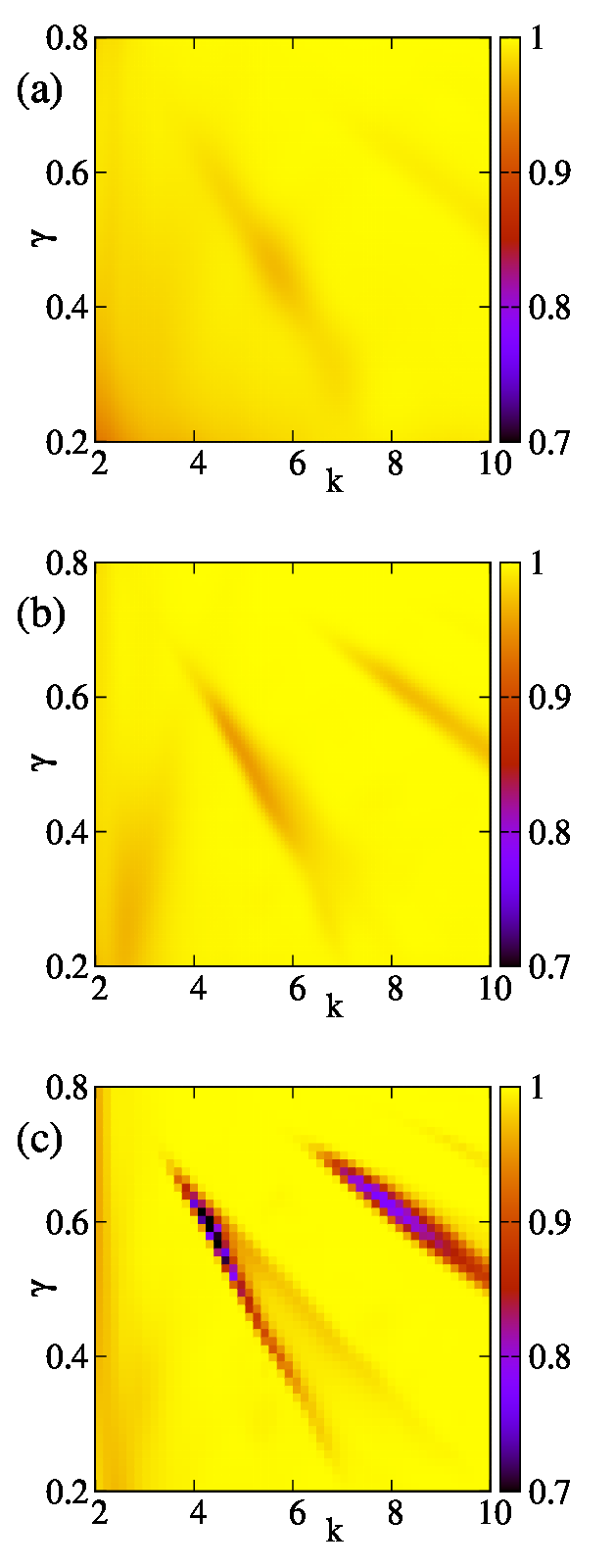} 
 \caption{(color online) Overlaps $O$ between the classical and quantum marginal distributions in parameter space $k,\gamma$. 
 Panel (a) shows $\hbar_{\rm eff}=0.412$, (b) $\hbar_{\rm eff}=0.137$, and (c) $\hbar_{\rm eff}=0.046$. 
 }
 \label{fig1}
\end{figure}

Aiming at detecting differences between the quantum evolution and the classical approximation given by 
our simple surmise, we evaluate the probability distributions $P(p)$ of momentum $p$ for both cases. 
In the classical case we take a discretized $P(p_i)$ distribution where the number of bins is given by the Hilbert 
space dimension used in the quantum calculations. 
These distributions correspond to the marginal ones associated to the Liouville and Wigner functions in the whole 
phase space, giving a faithful representation of their details despite the reduction in dimensionality and 
computational effort. The detailed study of the coherences which were 
initially considered in \cite{Carlo2}, will be the focus of future work. 
We evolve $10^6$ classical random initial conditions in the $p \in [-\pi;\pi]$ band of the 
cylindrical phase space, and an initial density matrix corresponding to these classical initial conditions 
for the quantum counterpart. We have evolved $5000$ classical time steps, and since the quantum equilibrium 
distribution is obtained in a few periods, we have taken just $50$ quantum ones.
As a first measure we calculate the overlaps $O=\int P_{\rm cl}(p) P_{\rm q}(p) dp$ over 
the whole parameter space. In Fig. \ref{fig1} we show the results for $\he=0.412$ in panel (a), 
$\he=0.137$ in panel (b), and $\he=0.046$ in panel (c) (we have taken a 100 by 100 points resolution in the first 
two cases, while 50 by 50 points in the third one). 
It is clear that this measure does not reveal significant differences between the classical and quantum distributions, 
almost all values are near $1$, with a few low values around $0.7$ for $\hbar_{\rm eff}=0.046$ essentially. 
If we look closer there is mainly a small area where the overlap is low and this difference 
keeps more or less the same location (but grows and enhances) with $\he \to 0$, roughly at the center left 
of the parameter space. There are two other disagreement areas, one is located around small $k$ 
and corresponds to very low forcing (whose overlaps are fairly large) and the other is on the upper right 
of the largest one (with lower overlaps but also enhancing for $\he \to 0$ ). We will refer to them later.

\begin{figure}[htp]
\includegraphics[width=0.47\textwidth]{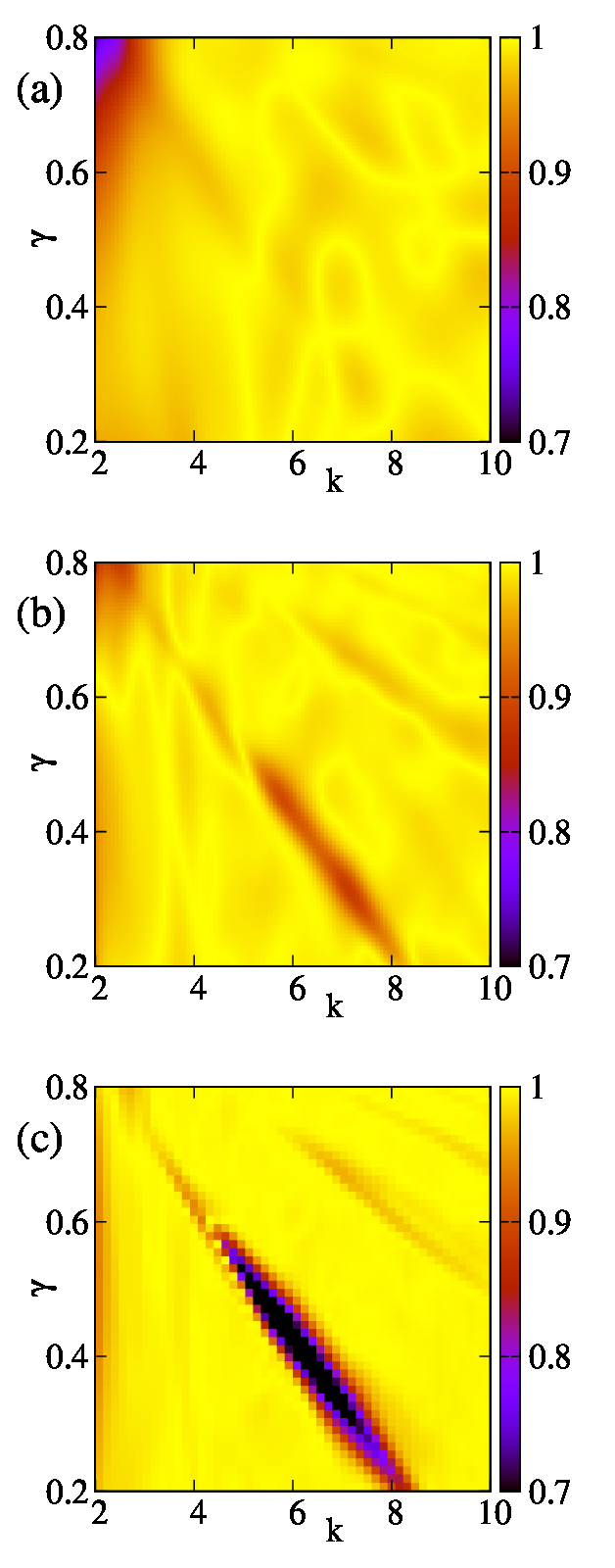} 
 \caption{(color online) Complement of the difference between the relative dispersions of the classical and quantum 
 marginal distributions $\sigma'$, in parameter space $k,\gamma$. Panel (a) shows $\hbar_{\rm eff}=0.412$, (b) $\hbar_{\rm eff}=0.137$, 
 and (c) $\hbar_{\rm eff}=0.046$. 
 }
 \label{fig2}
\end{figure}

In order to provide with another point of view for these deviations we calculate the 
complement of the difference between the relative dispersions of the classical and quantum $P(p)$ distributions given by  
$\sigma'=1-|\sigma_{\rm cl}-\sigma_{\rm q}|/(\sigma_{\rm cl}+\sigma_{\rm q})$. We take 
$\sigma=\sqrt{\langle p^2 \rangle - \langle p \ranglẹ^2}$ for 
both cases and $\langle \rangle$ stands for the corresponding statistical averages. This is displayed in Fig. \ref{fig2}, 
where we use the same three 
different $\he$ values of Fig. \ref{fig1} corresponding to panels (a), (b) and (c). 
It is clear that there is an overall agreement between the classical and 
quantum distributions. However there is again a main region with different behaviour that grows and enhances as $\he \to 0$. 
This time the location and shape is different from the one 
detected with the overlap measure, though sharing some area. Moreover, there are two smaller regions as before, 
located at a similar places and one in the upper left corner for the higher values of $\he$. In fact, this latter 
becomes the biggest one for $\he=0.412$ but it is significantly 
reduced when going towards the classical limit. Moreover, from $\sigma'$ point of view all this $\he$ case does not show a clear 
region of better or worse agreement with the exception of this region. This is due to the fact that fluctuations dominate 
this measure in this small sized basis, and amplify the differences for peaked distributions.

\begin{figure}[htp]
\includegraphics[width=0.47\textwidth]{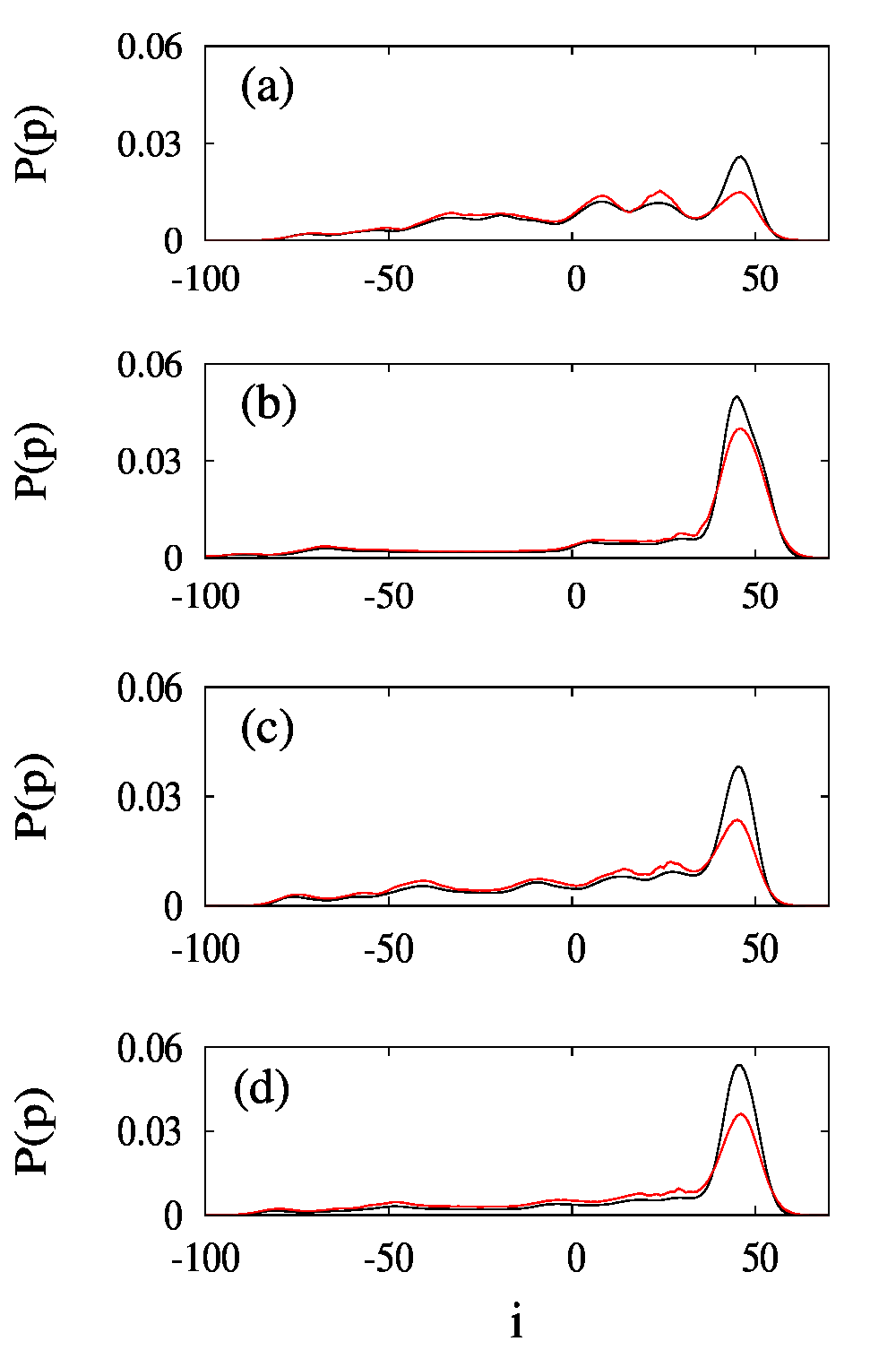} 
 \caption{(color online) Classical (black lines) and quantum ((red) gray lines) marginal distributions $P(p)$ as 
 a function of $i$, for 
 meaningful points in parameter space (see main text). In panel (a) $k=4.56$ and $\gamma=0.56$, in (b) $k=7.12$ and $\gamma=0.34$, 
 in (c) $k=5.2$ and $\gamma=0.48$, and in (d) $k=5.8$ and $\gamma=0.44$. In all cases $\hbar_{\rm eff}=0.137$.}
 \label{fig3}
\end{figure}

\begin{figure}[htp]
\includegraphics[width=0.47\textwidth]{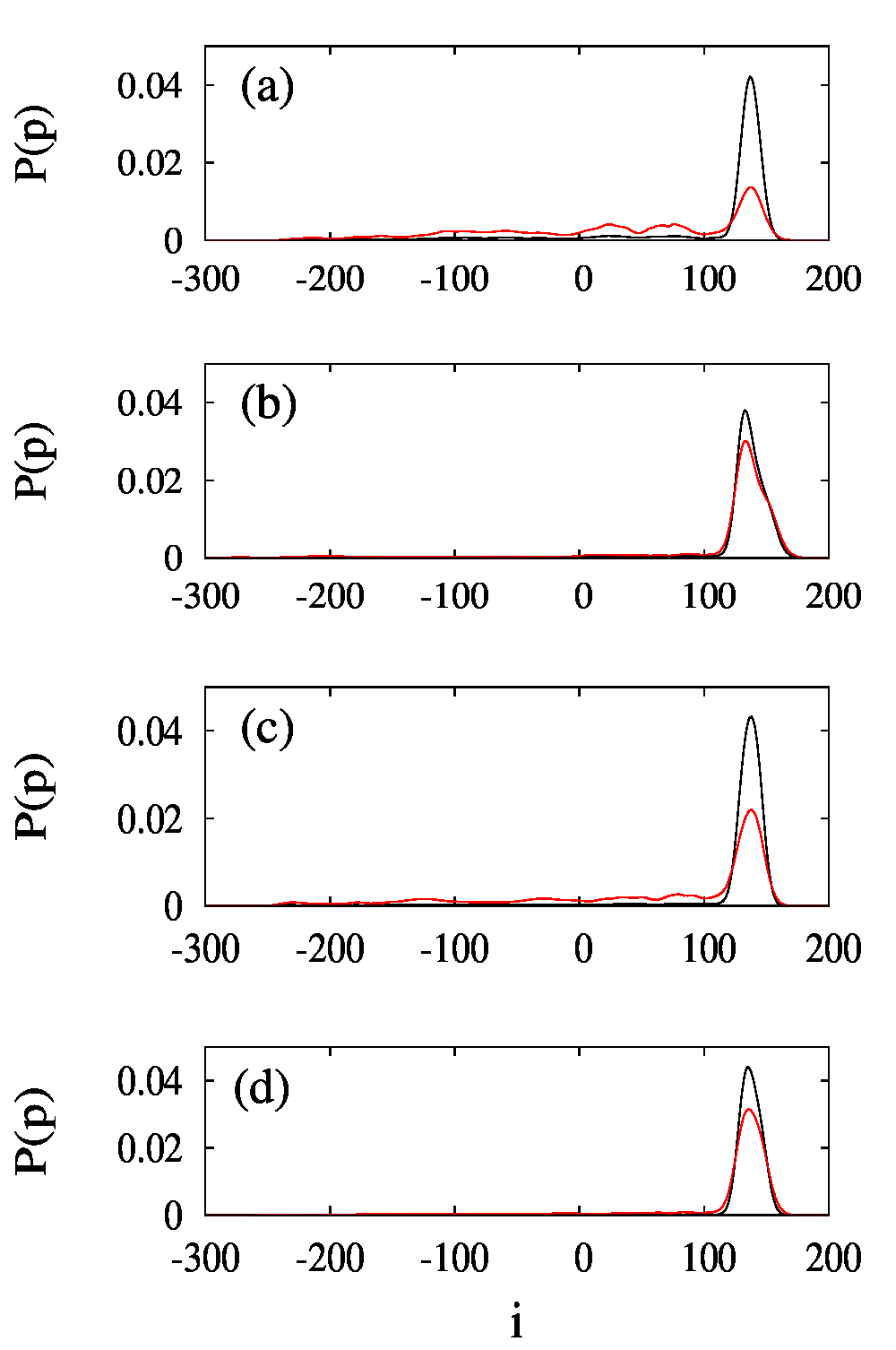} 
 \caption{(color online) Classical (black lines) and quantum ((red) gray lines) marginal distributions $P(p)$ as 
 a function of $i$, for 
 meaningful points in parameter space (see main text). In panel (a) $k=4.56$ and $\gamma=0.56$, in (b) $k=7.12$ and $\gamma=0.34$, 
 in (c) $k=5.2$ and $\gamma=0.48$, and in (d) $k=5.8$ and $\gamma=0.44$. In all cases $\hbar_{\rm eff}=0.046$.}
 \label{fig4}
\end{figure}

To clarify this behaviour, in Fig. \ref{fig3} we show the $P(p)$ distributions as a function of $i$, to also show the different 
dimension of the Hilbert space. We select four cases inside the largest difference regions for both measures. We 
begin with $\he=0.137$ where the area of interest is already defined for $\sigma'$. Fig. \ref{fig3}(a) roughly corresponds 
to the center of the minimum overlap region and Fig. \ref{fig3}(b) to the minimum $\sigma'$ region. In the latter case both 
distributions are more peaked than in the former. This makes small differences between the classical and quantum 
curves to enhance the distance in the dispersion measure more than in the overlap. On the contrary, when the distributions 
are more extended in $p$ the differences are better noticed through the overlap. 
To be exhaustive we show two more cases corresponding to the area of coincidence of both regions of maximum disagreement in overlap 
and $\sigma'$. These are shown in Fig. \ref{fig3}(c) for the overlap and 
in Fig. \ref{fig3}(d) for $\sigma'$. Again the same explanation as in the previous pair of cases is valid though 
the dispersion is a little larger. In Fig. \ref{fig4} we show the same results for $\he=0.046$. The behaviour is 
qualitatively the same though the Hilbert space dimension grows. This time it is clear that the overlap disagreement is larger 
(see Figs. \ref{fig4} (a) and (c)). For $\sigma'$ the maximum corresponds now to Fig. \ref{fig4} (d) and again this is due to a long 
quantum tail outside of the peak for negative $i$, that the classical model is not able to reproduce. 

\begin{figure}[htp]
\includegraphics[width=0.47\textwidth]{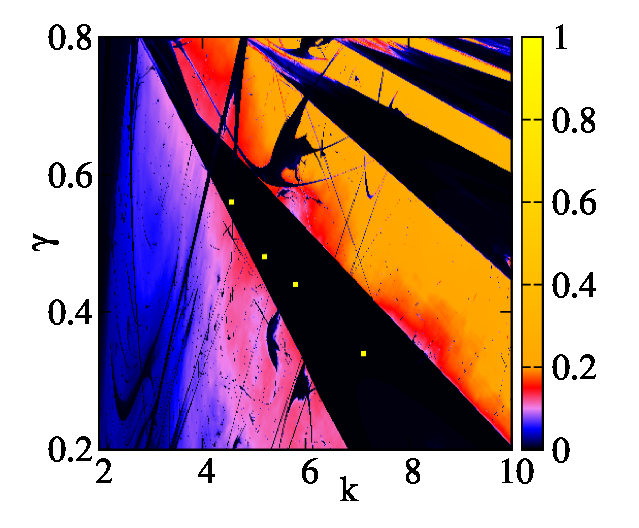} 
 \caption{(color online) Classical participation ratio $\eta$ in parameter space $k,\gamma$ without Gaussian noise. 
 We have taken $\hbar_{\rm eff}=0.137$. The four small (yellow) gray squares inside the largest regular region 
 correspond to the parameter values considered in Fig. \ref{fig3} and Fig. \ref{fig4}.}
 \label{fig5}
\end{figure}

But what are these two main regions that are different for these measures? 
As a matter of fact, they represent the same one. In Fig. \ref{fig5} we show a 
measure of chaoticity (or simplicity) of these distributions by means of 
the participation ratio $\eta=(\sum_iP(p_i)^2)^{-1}/N$. This measure has been originally defined as an indicator 
of the fraction of basis elements that expand a quantum state and we have extended this concept to the classical 
case by simply applying it to the discretized distribution.  
It is worth noticing that we explore the parameter space of the DMKRM in a meaningful big area where many regular 
isoperiodic stable structures (ISSs \cite{Celestino, Carlo, Ermann, Carlo2, Beims}, 
originally termed as periodicity hubs \cite{Gallas}) appear. 
They are characterized by low values of $\eta$ and 
can be noticed as the dark areas with sharp borders in Fig. \ref{fig5}.
We can see that both regions of disagreement between the classical and quantum distributions correspond 
to different subregions of the largest regular (black) one. 
Then, the difference has the same origin and this is the presence of the largest ISS in the 
DMKRM. Now, let's explain why the quantum subregions are different. Near the borders of this ISS 
the quantum asymptotic distributions 
explore a larger effective complex basin of attraction in addition to the simple limit cycle (of higher periodicity). 
In these cases the biggest difference is noticed by the overlap. When this basin is less explored by the 
asymptotic distributions the difference is more evident in the dispersion and this is associated to the inner 
part of the ISS. The enhancement as $\he \to 0$ reveals a compromise between the vanishing prefactor 
of quantum corrections to the classical evolution, and a better resolution of the larger regular regions. In fact, 
in the smaller regular domains the classical model behaves as in the chaotic background, while in the larger ones 
the correction dependence in $p$ could become relevant.

A word regarding the smaller areas of disagreement is in order here. 
The one appearing on the left hand side (i.e., at small $k$) 
of Fig. \ref{fig1} and Fig. \ref{fig2} corresponds to a big roughly regular region that is also associated to 
low forcing. The explanation for this behaviour is similar to 
the one for the big regular area, regularity enhances quantum classical differences. 
The small region approximately around $k=8$ and $\gamma=0.6$ belongs to the same family than the ISS previously studied, 
then again, the same reasoning applies.

This suggests the main result of our paper: chaos simplifies quantum friction. The terms that depend on the 
value of the momentum $p$ in the Wigner evolution equation (i.e. the dissipator given in Eq. (\ref{eq: DWlin})) become 
averaged out by the classical chaotic behaviour underlying the quantum counterpart. This is the main signature of 
chaos in this kind of quantum dissipative systems. The dependence in $\gamma$ is partly embedded in the dynamics after 
each map iteration. The last classical map step that has a $\gamma$ independent 
Gaussian noise simply stands for the $\he$ size quantum uncertainty that is embodied in Eq. (\ref{qcoarse}).

\section{Conclusions}
\label{sec4}

We have derived the evolution equation for the Wigner distributions of dissipative systems. In particular 
we have evaluated the case with friction using the standard model found in the literature and 
employed by us in previous work. At first sight, it turns out that the quantum corrections to the 
classical evolution of Liouville distributions depend non-trivially on the phase space variable $p$ and 
the dissipation strength $\gamma$ (i.e. the coupling with the environment). By further studying a 
paradigmatic model of quantum dissipative systems given by a modified kicked rotator (namely the DMKRM \cite{qdisratchets}) 
we confirm that the main features of the asymptotic Wigner distributions can be obtained by a very simple classical Gaussian noise 
model. The noise strength is solely given by the size of $\hbar$. In fact, when we systematically analyze 
the overlap of the marginal distributions of the Wigner functions in the whole parameter space, we find that 
the agreement is uniform with the exception of the largest regular regions. Moreover, when looking at the 
dispersion of these distributions we find a complementary behaviour. The difference is related to the specific 
shape of the quantum distributions in different subregions inside these largest regular domains, and helps us to identify 
a shared source for them. 

We conjecture that the underlying chaotic dynamics is responsible for self averaging the quantum corrections terms containing 
$p$, erasing the memory of this dependence at each time step of the map. A strong hint on this is given by the fact that the 
only deviations between the classical and quantum results are located at the largest regular regions of the whole 
parameter space. The dependence on $\gamma$ is present through the dynamics, in fact the noise of 
the previous steps of the map suffer form a $\gamma$ sized contraction. Finally, the last step of the noise is related 
to the $\hbar$ sized quantum uncertainty embodied in the first terms of the quantum corrections to the classical 
evolution. In the future, we plan to rigorously proof this conjecture by means of a simplified model.

\section*{Acknowledgments}

Support from CONICET is gratefully acknowledged. 

\vspace{3pc}



\begin{thebibliography}{99}

\bibitem{SaracenoPaz} 
P. Bianucci, J.P. Paz, and M. Saraceno, 
Phys. Rev. E {\bf 65}, 046226 (2002);
L. Ermann, J.P. Paz, and M. Saraceno, 
Phys. Rev. A {\bf 73}, 012302 (2006).

\bibitem{DelCampo}
Z. Xu, L.P. García-Pintos, A. Chenu, and A. del Campo, 
arXiv:1810.02319.

\bibitem{OTOC}
A. Lakshminarayan, 
arXiv:1810.12019; 
E.B. Rozenbaum, S. Ganeshan, and V. Galitski, 
Phys. Rev. Lett {\bf 118}, 086801 (2017);
I. García-Mata, M. Saraceno, R.A. Jalabert, A.J. Roncaglia, and D.A. Wisniacki, 
arXiv:1806.04281.

\bibitem{qfield}
J. Maldacena, S.H. Shenker, and D. Stanford, 
JHEP {\bf 2016}, 106 (2016).

\bibitem{Bakemeier}
L. Bakemeier, A. Alvermann, and H. Fehske, 
Phys. Rev. Lett. {\bf 114}, 013601 (2015).

\bibitem{Hartmann}
M. Hartmann, D. Poletti, M. Ivanchenko, S. Denisov, and P. H\"anggi, 
New J. Phys. {\bf 19}, 083011 (2017).

\bibitem{Ivanchenko}
M. Ivanchenko, E. Kozinov, V. Volokitin, A. Liniov, I. Meyerov, and S. Denisov, 
Ann. Phys. (Berlin) {\bf 529}, 1600402 (2017); 
I.I. Yusipov, O.S. Vershinina, S.V. Denisov, S.P. Kuznetsov, and M.V. Ivanchenko, 
arXiv:1806.09295.

\bibitem{Wang}
R.R.W. Wang, B. Xing, G.G. Carlo, and D. Poletti, 
Phys. Rev. E {\bf 97}, 020202(R) (2018).

\bibitem{Eastman}
J.K. Eastman, J.J. Hope, and A.R.R. Carvalho, 
Sci. Rep. {\bf 7}, 44684 (2017).

\bibitem{Eastman2}
J.K. Eastman,  S.S. Szigeti, J.J. Hope, and A.R.R. Carvalho, 
arXiv:1808.09257.

\bibitem{Bondaretal} D. I. Bondar, R. Cabrera, A. Campos, S. Mukamel, and H. A. Rabitz, 
\emph{J. Phys. Chem. Lett.} \textbf{7}, 1632 (2016).

\bibitem{qdisratchets}
G.G. Carlo, G. Benenti, G. Casati, and D.L. Shepelyansky,
Phys. Rev. Lett. {\bf 94}, 164101 (2005).

\bibitem{Curtright1998}
T. Curtright, D. Fairlie, and C. Zachos,
Phys. Rev. D \textbf{58}, 025002 (1998).

\bibitem{Curtright2011}
T.L. Curtright and C.K. Zachos, 
Asia Pacific Physics Newsletter {\bf 1}, 37 (2012).

\bibitem{Curtright2013}
T. Curtright, D.B. Fairlie, and C.K. Zachos, 
\emph{A Concise Treatise on Quantum Mechanics in Phase Space} 
(World Scientific, Singapore, 2013).

\bibitem{Gardiner1985}
C. Gardiner,
\emph{Stochastic methods} 
(Springer, Berlin, 1985).

\bibitem{Dittrich}
T. Dittrich and R. Graham,
Europhys. Lett. {\bf 7}, 287 (1988).

\bibitem{Graham}
R. Graham,
Z. Phys. B Cond. Mat. {\bf 59}, 75 (1985).

\bibitem{opetor} A.M.F. Rivas and A.M. Ozorio de Almeida, 
Ann. Phys. (N. Y.) \textbf{276}, 223 (1999).

\bibitem{Carlo}
G.G. Carlo,
Phys. Rev. Lett. {\bf 108}, 210605 (2012).

\bibitem{Lindblad}
G. Lindblad,
Commun. Math. Phys. {\bf 48}, 119 (1976).

\bibitem{Carlo2}
G.G. Carlo, A.M.F. Rivas, and M.E. Spina, 
Phys. Rev. E {\bf 92}, 052907 (2015);
G.G. Carlo, L. Ermann, A.M.F. Rivas, and M.E. Spina, 
Phys. Rev. E {\bf 93}, 042133 (2016).

\bibitem{Celestino}
A. Celestino, C. Manchein, H.A. Albuquerque, and M.W. Beims,
Phys. Rev. Lett. {\bf 106}, 234101 (2011).

\bibitem{Ermann}
L. Ermann and G.G. Carlo,
Phys. Rev. E {\bf 91}, 010903(R) (2015).

\bibitem{Beims}
M.W. Beims, M. Schlesinger, C. Manchein, A. Celestino, A. Pernice, and W.T. Strunz,
Phys. Rev. E {\bf 91}, 052908 (2015).

\bibitem{Gallas}
J.A.C. Gallas, 
Phys. Rev. Lett. {\bf 93}, 2714 (1993).

\end{thebibliography}
\end{document}